\documentclass[prb,showpacs,aps,twocolumn]{revtex4} %preprint
\usepackage{amsmath}
\usepackage{epsfig}

\begin{document}
\title{High-field magneto-excitons in unstrained GaAs/Al$_{x}$Ga$_{1-x}$As quantum dots}
\author{Y. Sidor}
\email{yosyp.sidor@ua.ac.be}
\author{B. Partoens}
\author{F. M. Peeters}
\email{francois.peeters@ua.ac.be} \affiliation{Departement Fysica,
Universiteit Antwerpen (CGB), Groenenborgerlaan 171, B-2020
Antwerpen, Belgium}
\author{N. Schildermans, M. Hayne and V. V. Moshchalkov} \affiliation{Pulsed
Field Group, Laboratory of Solid State Physics and Magnetism,
K.U.Leuven, Celestijnenlaan 200D, B-3001 Leuven, Belgium}
\author{A. Rastelli, and O. G. Schmidt} \affiliation{Max-Planck-Institut
f\"{u}r Festk\"{o}rperforschung, Heisenbergstrasse 1, D-70569
Stuttgart, Germany}
\date{\today}

\begin{abstract}

The magnetic field dependence of the excitonic states in
unstrained GaAs/Al$_{x}$Ga$_{1-x}$As quantum dots is investigated
theoretically and experimentally. The diamagnetic shift for the
ground and the excited states are studied in magnetic fields of
varying orientation. In the theoretical study, calculations are
performed within the single band effective mass approximation,
including band nonparabolicity, the full experimental
three-dimensional dot shape and the electron-hole Coulomb
interaction. These calculations are compared with the experimental
results for both the ground and the excited states in fields up to
50 Tesla. Good agreement is found between theory and experiment.

\end{abstract}

\pacs{73.21.La, 75.75.+a, 78.67.Hc}

\maketitle

\section{Introduction}

Excitonic properties in quantum nanostructures have been
intensively investigated over the past decade because of the high
potential for applications and fundamental
physics~\cite{Apl1,Apl2,Apl3}. Among these nanostructures, quantum
dots (QDs) are three-dimensional wells that can trap electrons and
holes resulting in quantized energy levels. The density of states
is $\delta$ function-like, and the wave function of the particles
is localized inside the dot. Therefore, excitons play an important
role for the optical properties in such QDs. To explore these QD
properties, optical spectroscopy experiments in the presence of an
external magnetic field is a good probing
tool~\cite{ExpMagB1,Leuven}.

Semiconductor QDs can be grown by many methods, but in most cases
are based on lithography and self-organization techniques.
Self-assembled QDs formed through the Stranski-Krastanow growth
mode, among the various 0D systems investigated so far, are the
most well studied QDs. They can be fabricated with high uniformity
and in a single layer (or in arrays of vertically aligned stacks).
The optical properties of self-assembled QDs in the presence of an
external magnetic field have been studied theoretically and
experimentally by several groups. Early experimental work on InAs
self-assembled QDs by Zhu \emph{et. al.}~\cite{ExpArbB1} and also
by several other groups~\cite{ExpB1,ExpB2,ExpB3,ExpB4} observed
the excitonic properties of different self-assembled dots in the
presence of parallel and perpendicular magnetic fields. One should
mention the interesting experimental and theoretical work done by
Bayer \emph{et. al.}~\cite{ExpArbB2}, where self-assembled dots
were investigated in magnetic fields of varying orientations.
Theoretical approaches, based on a harmonic oscillator confinement
potential and perturbation theory (see for example
Ref.~\onlinecite{ThGarmOsc2}) have been presented in order to
calculate the diamagnetic shift of the exciton ground state energy
and to compare with the experimental data. Excited states in the
dot were studied previously (see for example
Refs.~\onlinecite{Ref2,ExciteB1a,ExciteB2a,Karen}), both
experimentally and theoretically by Raymond \emph{et
al.}~\cite{ExciteB1} and in another experimental work by Pulizzi
\emph{et al.}~\cite{ExciteB2}. In the work by Raymond \emph{et
al.} valence band mixing was included within the eight-band
\textbf{k}$\cdot$\textbf{p} model, coupled via the Bir-Pikus
Hamiltonian to the strain of the dot calculated using classical
elasticity theory. Very recently, experimental and theoretical
results within the single band effective mass approximation
including strain were presented in the absence of an external
magnetic field, where the dot structure was taken directly from
the experiment (possibly full 3D calculations). The ground state
photoluminescence energy of the InAs/AlAs and InAs/GaAs QDs was
calculated~\cite{SingleBand}.

GaAs/AlGaAs quantum dots cannot be created by Stranski-Krastanov
growth because of the almost perfect match of lattice parameters.
Nevertheless, this system offers several advantages compared to
others: the grown material is ideally unstrained, and sharp
interfaces with reduced intermixing can be obtained; they can be
designed to emit light in the optimum spectral range of sensitive
Si-based detectors. A new experimental method was presented
recently to obtain GaAs/AlGaAs quantum dots via multistep
(hierarchical) self-assembly. First a template of InAs/GaAs
islands was created, which are "converted" into nanoholes on a
GaAs surface by GaAs overgrowth followed by in situ etching.
Self-assembled nanoholes are then transferred to an
Al$_{x}$Ga$_{1-x}$As surface, filled with GaAs, and overgrown with
Al$_{y}$Ga$_{1-y}$As (x$>$y). In this case the quantum dot
morphology is given by the shape of the Al$_{x}$Ga$_{1-x}$As
nanoholes~\cite{ExpGaAsAlGaAs}. This allows the electronic
properties of the QDs to be calculated without the complication of
uncertain composition and strain profiles~\cite{ExpGaAsAlGaAs,
Rastelli1,Rastelli2}. The GaAs QDs are characterized by widely
tunable emission energy, narrow inhomogeneous broadening (of the
order of 8-20 meV depending on the growth parameters) and large
separation between QD emission and continuum (up to $\sim$200
meV). Single-QD spectroscopy on such QDs show resolution-limited
sharp lines~\cite{ExpGaAsAlGaAs,Rastelli3,Rastelli4},
single-photon emission up to 77 K and bright emission up to room
temperature~\cite{Rastelli4}.

In this paper we focus our attention on the energy spectrum of
single excitons in the presence of an applied magnetic field for
such unstrained GaAs/Al$_{x}$Ga$_{1-x}$As
QDs~\cite{ExpGaAsAlGaAs}. We perform calculations within the
single band effective mass approximation but including band
nonparabolicity in first order approximation. The real
experimental shape of the dot with a quantum well on the top and
asymmetrical AlGaAs barriers is considered. The calculations are
based on a full 3D finite element scheme. We study the exciton
energy shift for the ground and the excited states of the QD in
magnetic fields up to 50 T of varying orientation.

The paper is organized in the following way. In Sec. II we briefly
describe the method and the QD model used in our calculations. In
Sec. III we discuss the Coulomb interaction energies and the wave
function extensions of the particles in the presence of a magnetic
field. Finally in Sec. IV, we present results of the influence of
a magnetic field on the excitonic spectrum in
GaAs/Al$_{x}$Ga$_{1-x}$As QDs, comparing theoretical and
experimental data.

\section{Theoretical formalism and model}

The geometry of the quantum dot was taken directly from scanning
tunnelling microscopy (STM) measurements. The profile of the 3D
dot is illustrated in Fig. 1. On top of the GaAs dot-well system,
an Al$_{0.35}$Ga$_{0.65}$As barrier was deposited, while on the
bottom Al$_{0.45}$Ga$_{0.55}$As. Therefore, this structure is
slightly non-symmetric due to the different barriers. Fig. 1 also
shows the height ($h$) distribution in the 2D lateral profile of
the dot with the base widths $w_{x}$ and $w_{y}$. According to it,
the dot is more elongated along the $y$-direction, and the height
distribution is not uniform with respect to the center of the
$xy$-plane.

The conduction and the valence bands are centered around the
$\Gamma$-valley of GaAs and Al$_{x}$Ga$_{1-x}$As. Within the
single band effective-mass theory, in order to describe electron
and hole states of the dots, we solve the Schr\"{o}dinger
equations

\begin{equation}
H\Psi(\textbf{r}_{\textbf{e}},\textbf{r}_{\textbf{h}})=E\Psi(\textbf{r}_{\textbf{e}},\textbf{r}_{\textbf{h}}),
\end{equation}
with the Hamiltonian

\begin{equation}
H =H_{e}+H_{h}+U(\textbf{r}_{\textbf{e}}-\textbf{r}_{\textbf{h}}),
\end{equation}
where $H_{e(h)}$ denotes the single electron (hole) Hamiltonian,
$U(\textbf{r}_{\textbf{e}}-\textbf{r}_{\textbf{h}})$ is the
Coulomb interaction between the electron and the hole
\begin{equation}
U(\textbf{r}_{\textbf{e}}-\textbf{r}_{\textbf{h}})=-\frac{e^{2}}{\varepsilon
|\textbf{r}_{\textbf{e}}-\textbf{r}_{\textbf{h}}|} ,
\end{equation}
where \emph{e} is the charge of the electron, and $\varepsilon$ is
the static dielectric constant taken as the value of GaAs, because
of the very small difference of $\varepsilon$ inside and outside
the dot (see Table I).

\begin{figure} [t]
\vspace{0.4 cm} \ \epsfig{file=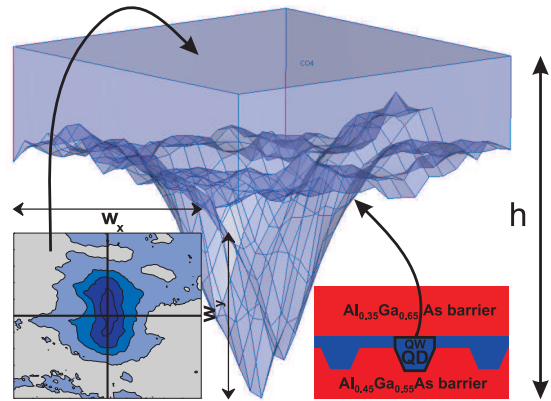,width=8 cm} \caption{
(color online). Geometry of the GaAs/Al$_{x}$Ga$_{1-x}$As QD, as
measured by STM topography. The height $h$ = 6.45 nm and base
widths along $x-$ and $y-$directions $w_{x}$ = 56 nm and $w_{y}$ =
72 nm, respectively ($x-$direction corresponds to the [1-10]
crystal direction of GaAs). On the top of the dot a thin QW layer
is deposited (for details see Refs.~\onlinecite{ExpGaAsAlGaAs}).}
\end{figure}

In the presence of a magnetic field the momentum is
replaced by the expression \textbf{p} $\rightarrow$
$\textbf{p}-\frac{q}{c}\textbf{A}$, so that the kinetic operator
for the particles becomes
\begin{eqnarray}
T=\left(\textbf{p}-\frac{q}{c}\textbf{A}\right)\frac{1}{2m^{\ast}}\left(
\textbf{p}-\frac{q}{c}\textbf{A}\right),
\end{eqnarray}
where $q$ is the charge of the particle, $m^{\ast}$ the spatial
dependent mass of the particle and \textbf{A} is the vector
potential. Let us consider the case, when the magnetic field is
applied along an arbitrary direction. We choose the symmetric
gauge

\begin{eqnarray}
\lefteqn{ \textbf{A} = -\frac{1}{2} \textbf{r} \times \textbf{B} }
\nonumber\\
& & = \frac{1}{2}((B_{y}z - B_{z}y)\hat{x}+(B_{z}x -
B_{x}z)\hat{y}\nonumber +(B_{x}y - B_{y}x)\hat{z}),
\end{eqnarray}
where $B_{x}$, $B_{y}$ and $B_{z}$ are the components  of the
magnetic field along $x$-, $y$- and $z$-directions, respectively,
given by

\begin{eqnarray}
\lefteqn{ B = \left(B_{x},B_{y},B_{z}\right) }
\nonumber\\
& & =
B\left(\cos(\varphi)\sin(\theta),\sin(\varphi)\sin(\theta),\cos(\theta)\right),
\end{eqnarray}
where (0 $\leq \theta \leq \pi$, 0 $\leq \varphi \leq$ 2$\pi$).
Inclusion of the magnetic field leads to fifteen additional
diamagnetic terms in the kinetic term, in addition to
$\textbf{p}(1/(2m^{\ast}))\textbf{p}$. From them, twelve are
linear with respect to $B_{i}$, $r_{i}$ and $\nabla_{i}$ (i = x, y
and z; here and further we will use $r_{x}$ = $x$, $r_{y}$ = $y$
and $r_{z}$ = $z$), and three other are quadratic

\begin{eqnarray}
C\sum_{i(j,k)=x}^{y,z}\frac{(B_{i}r_{j} -
B_{j}r_{i})^{2}}{m^{\ast}_{k}},
\end{eqnarray}
where the constant $C = q^{2}/(8c^{2})$, and $m^{\ast}_{k}$ are
the position dependent components of the effective mass tensor.

In the QD structure the subband energy due to the
\emph{z-}confinement is typically of the order of 100 meV.
Therefore, the anisotropy and the corrections due to the
conduction band nonparabolicity are
important~\cite{SchildNonparab}. The electron Hamiltonian is

\begin{eqnarray}
&&H_{e}(r_{e})=-\nabla_{xe}\frac{\hbar^{2}}{2m^{\ast}_{\parallel
e}(\textbf{r}_{e})}\nabla_{xe}
-\nabla_{ye}\frac{\hbar^{2}}{2m^{\ast}_{\parallel e}(\textbf{r}_{e})}\nabla_{ye}\nonumber\\
&&-\nabla_{ze}\frac{\hbar^{2}}{2m^{\ast}_{\perp
e}(\textbf{r}_{e})}\nabla_{ze}+V_{e}(\textbf{r}_{e})+V_{eB}(\textbf{r}_{e}),
\end{eqnarray}
where $V_{e}(r_{e})$ is the conduction band offset and we took the
asymmetry of the AlGaAs barriers and the geometry of the dot into
account (see Fig. 1). $V_{eB}(r_{e})$ denotes the aforementioned
electron diamagnetic terms, $m^{\ast}_{\parallel e}$ and
$m^{\ast}_{\perp e}$ are the perpendicular and parallel effective
masses of the electron, which we obtain from the first order
nonparabolicity approximation~\cite{nonparab}

\begin{subequations}
\begin{eqnarray}
m^{\ast}_{\perp e} = m^{\ast}_{e}(1+\alpha E),
\end{eqnarray}
\begin{eqnarray}
m^{\ast}_{\parallel e} = m^{\ast}_{e}(1+(2\alpha+\beta)E),
\end{eqnarray}
\end{subequations}
where $m^{\ast}_{e}$ is the bulk electron masses of the material,
$\alpha$ and $\beta$ are the nonparabolicity parameters (see Table
I), and E is the ground state energy of the electron obtained by
solving the single particle Schr\"{o}dinger equation using the
bulk masses of the electron. The parallel mass determines the
electron energy in the $xy$-plane, and the perpendicular mass
determines the quantization energy of the electrons in the
$z$-direction.

We consider both the heavy- (hh) and the light- (lh) hole states
in our calculation. The strain is not included because the QD
structure is unstained due to the experimental growth
conditions~\cite{ExpGaAsAlGaAs}. The single particle Hamiltonian
for both holes (h) is

\begin{eqnarray}
&&H_{h}(r_{h})=-\nabla_{xh}\frac{\hbar^{2}}{2m^{\ast}_{
h}(\textbf{r}_{h})}\nabla_{xh}
-\nabla_{yh}\frac{\hbar^{2}}{2m^{\ast}_{
h}(\textbf{r}_{h})}\nabla_{yh}\nonumber\\
&&-\nabla_{zh}\frac{\hbar^{2}}{2m^{\ast}_{
h}(\textbf{r}_{h})}\nabla_{zh}+V_{h}(\textbf{r}_{h})+V_{hB}(\textbf{r}_{h}),
\end{eqnarray}
where $m^{\ast}_{h}$ is the effective masses of the hole,
$V_{h}(r_{h})$ is the valence band confinement offset of the hole,
$V_{hB}(r_{h})$ denotes the hole diamagnetic terms. Note that this
single particle Hamiltonian is not just the diagonal part of the
four-band \textbf{k}$\cdot$\textbf{p} Hamiltonian. The mass in
each direction is the same and corresponds to the curvature of the
hole bands around the $\Gamma$ point, \emph{i.e.}, for the
heavy-hole and the light-hole, respectively

\begin{subequations}
\begin{eqnarray}
\frac{m_{0}}{m^{\ast}_{ hh}} = \gamma_{1}-2\gamma_{2},
\end{eqnarray}
\begin{eqnarray}
\frac{}{}\frac{}{}\frac{}{} \frac{m_{0}}{m^{\ast}_{ lh}} =
\gamma_{1}+2\gamma_{2},
\end{eqnarray}
\end{subequations}
where $\gamma_{1}$ and $\gamma_{1}$ are Luttinger parameters (see
Table I), and $m_{0}$ is the vacuum electron mass.

\begin{table}
\caption{\label{tab:table1} Material parameters used in the
calculations: band gap $E_{g}$, electron mass $m_{e}$
(Ref.~\onlinecite{AlGaAsElmass}), Luttinger parameters
$\gamma_{1}$ and $\gamma_{1}$, nonparabolicity parameters $\alpha$
and $\beta$ (Ref. ~\onlinecite{nonparab}), and dielectric constant
$\varepsilon$ (Ref.~\onlinecite{BookParameter}). For
Al$_{x}$Ga$_{1-x}$As material parameters, the first-order
interpolation formula is used: parameter(Al$_{x}$Ga$_{1-x}$As) =
parameter(GaAs) $\pm$ $x$(parameter(GaAs)-parameter(AlAs)).}
\begin{ruledtabular}
\begin{tabular}{cccccc}
Parameter&GaAs&Al$_{0.35}$Ga$_{0.65}$As&Al$_{0.45}$Ga$_{0.55}$As \\
\hline
$E_{g}$(eV) & 1.519 & 1.956 & 2.080 &  \\
$m_{e}$(m$_{0}$) & 0.067 & 0.096 & 0.104 & \\
$\gamma_{1}$ & 6.98 & 5.85 & 5.53 & \\
$\gamma_{2}$ & 2.06 & 1.63 & 1.50 & \\
$\alpha$(eV$^{-1}$) & 0.64 & -- & -- & \\
$\beta$(eV$^{-1}$) & 0.70 & -- & -- & \\
$\varepsilon$ & 12.9 & 11.9 & 11.6 & \\
\end{tabular}
\end{ruledtabular}
\end{table}

To include the Coulomb interaction between the electron and the
hole in the QD, we calculate the Hartree potential $\varphi(r)$ by
solving the 3D Poisson equation:

\begin{eqnarray}
\varepsilon\nabla^{2}\varphi(\textbf{r})=-4\pi\rho(\textbf{r}),
\end{eqnarray}
where $\rho(\textbf{r})$ is the particle density, previously
calculated from Eq. (1). The potential $\varphi(\textbf{r})$ is
the one felt by the hole, \emph{i.e.}, in mathematical terms it
means $\varphi(\textbf{r}_{h}) \approx \int
d\textbf{r}_{e}\rho(r_{e})/|\textbf{r}_{e}-\textbf{r}_{h}|$. To
solve the Poisson equation, often one uses zero as the boundary
condition, which is a good approximation when a large enough grid
is used. The regular asymptotic boundary condition to the Poisson
equation is proportional to $1/r$. Note, because the dot profile
is not symmetric, consequently the density of the electron is not
symmetric, and therefore the exact asymptotic boundary condition
for the last equation becomes
\begin{eqnarray}
\varphi(\textbf{r}_{e})=\frac{q}{\varepsilon|\textbf{r}_{e}-\textbf{r}_{0
e}|},
\end{eqnarray}
where $\textbf{r}_{0 e}=\int
\textbf{r}_{e}\rho(\textbf{r}_{e})d\textbf{r}_{e}$ is the average
value of the coordinates of the electron, and \emph{q} is the
electron charge. The Coulomb energy, which accounts for the
interaction between the electron-hole pair, is obtained by
performing 3D integration over the hole coordinates

\begin{eqnarray}
E_{C}=\int \varphi(\textbf{r}_{h})\rho(\textbf{r}_{h})d
\textbf{r}_{h},
\end{eqnarray}
where $\rho(r_{h})$ is the hole density. Note, that by solving
the Poisson equation we circumvent the direct 6D integration over
electron and hole space, and we avoid the 1/r singularity problem
for r $\rightarrow$ 0.

\begin{figure}
\vspace{-0.5cm} \ \epsfig{file=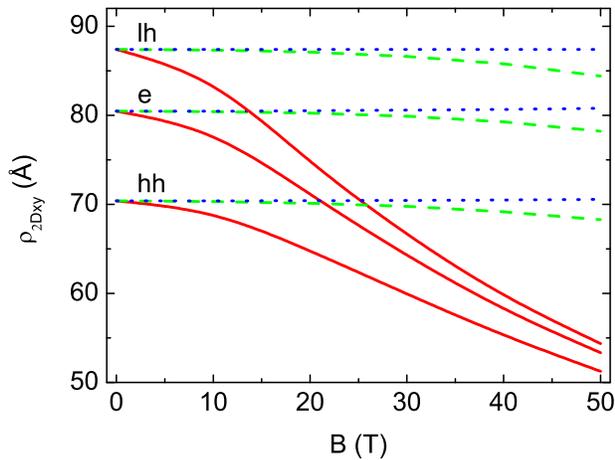,width=9.5cm}
\vspace{-0.8cm} \ \caption{(color online). The extent of the
electron (e), the heavy-hole (hh) and the light-hole (lh) in the
\emph{xy}-plane, as a function of the magnetic field \textbf{B}
for GaAs/Al$_{x}$Ga$_{1-x}$As QD. The dashed (green), dotted
(blue), and full (red) curves correspond to the radii, when the
magnetic field is applied along the \emph{x-}, \emph{y-}, and
\emph{z-}direction, respectively.}
\end{figure}

The full 3D calculations with the real shape of the dot in the
presence of the magnetic field, as well as the Poisson equation
are performed using the finite element method on a variable size
grid. The input parameters used in the simulations are presented
in Table I. Most of the parameters have been taken from
Ref.~\onlinecite{Parameter}, otherwise indicated.

\section{Extent of the particles and the Coulomb interaction energies}

In this section we discuss the electron-hole interaction and the
wave function extensions of the particles in the presence of a
magnetic field. \emph{One should stress, that there are no fitting
parameters in all our calculations}.

\begin{figure}
\vspace{-0.5cm} \ \epsfig{file=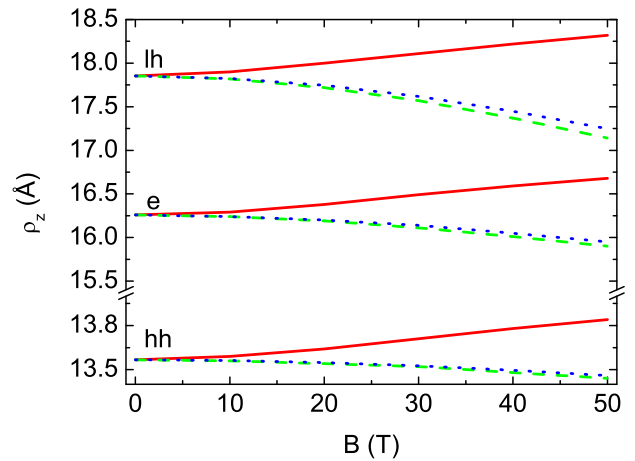,width=9.5cm}
\vspace{-0.8cm} \ \caption{(color online). The same as Fig. 2 but
now in the \emph{z-}direction.}
\end{figure}

The mass of the heavy-hole inside the dot, is equal to
0.35$m_{0}$, and is four times larger than the mass of the
light-hole 0.09$m_{0}$ (see Eqs. (10a), (10b) and Table I). As a
result, the wave function radii in the \emph{xy}-plane of the
heavy-hole for zero magnetic field \textbf{B} are smaller than for
the light-hole, as can be seen from Fig. 2. However, one can
notice that the electron extent is smaller than for the
light-hole. The reason is that, although the electron mass is a
bit lighter than the one of the light-hole, the conduction band
offset is roughly 1.5 times larger than the offset for the
light-hole. Let us comment on the behavior of the particle radii
in the case when the magnetic field is applied along the
\emph{z-}direction. The magnetic field influences the particle
motion in the plane perpendicular to the applied direction. When
the magnetic field is applied along the \emph{z-}direction
(influence on the \emph{xy}-plane), the 2D radii in the
\emph{xy}-plane are strongly squeezed for all particles (see full
curves in Fig. 2), and it leads to a decrease of at least 30 \%
(even more for the light-hole). This large change is due to the
flatness of the dot, with a lateral dimension about 10 times
larger than its height. This is the reason why the magnetic field
applied along the \emph{x-} and \emph{y-}directions has a less
pronounced effect on the electron and holes 2D extension (see
dashed and dotted curves in Fig. 2, respectively). The dot is
slightly larger in the \emph{y-}direction which explains the small
difference between the dashed and dotted curves in Fig. 2.

Radii along the \emph{z-}direction are significantly smaller than
those in the plane, as depicted in Fig. 3. The electron radius is
smaller than the one for the light-hole, because of the difference
of conduction and valence band offsets. When the magnetic field is
applied along the \emph{z-}direction, the wave function is
squeezed in the plane of the dot, as was already observed, but at
the same time it will become elongated in the \emph{z-}direction
(full curves in Fig. 3), because the total particle probability is
a constant. This is similar to squeezing a balloon in two
directions. In the case of the magnetic field applied along the
\emph{x-} and \emph{y-}directions (dashed and dotted curves in
Fig. 3, respectively), the particle radii slightly decrease with
increasing magnetic field due to the squeezing of the wave
function in, respectively, the \emph{yz-} and \emph{xz-} plane.

\begin{figure}
\vspace{-0.5cm} \ \epsfig{file=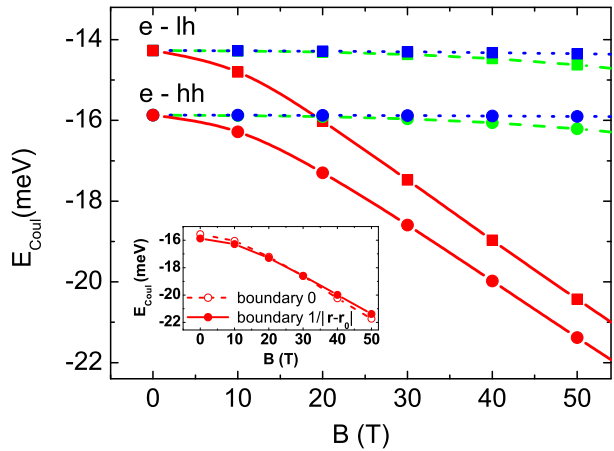,width=9.5cm}
\vspace{-0.8cm} \ \caption{(color online). The electron-hole
Coulomb interaction energy as a function of magnetic field for
GaAs/Al$_{x}$Ga$_{1-x}$As QD. The dashed (green), dotted (blue),
and full (red) curves with the circles correspond to the
heavy-hole-electron Coulomb energies, for the magnetic field
applied along the \emph{x-}, \emph{y-}, and \emph{z-}direction,
respectively. The curves with the squares represent the
light-hole-electron Coulomb interaction energies. The inset shows
the Coulomb energy, as a function of magnetic field applied along
the \emph{z-}direction. The dashed (full) curves with the open
(closed) circles correspond to the two different boundary
conditions of the Poisson equation.}
\end{figure}

The dependence of the electron-hole Coulomb interaction on the
external magnetic field is shown in Fig. 4. As was mentioned in
Sect. II, one can use different boundary conditions for the
Poisson equation. In the inset of Fig. 4 we show that by using the
boundary condition described by Eq. (12) and the one with zero
boundary conditions may influence the Coulomb interaction energy
slightly. The absolute value of the Coulomb interaction is
inversely proportional to the relative distance between the
electron and the hole. The radii for all the particles in the
\emph{xy}-plane are much larger than that along the
\emph{z-}direction. Therefore, the interaction between the
electron and the hole is predominantly determined by the behavior
of the particle radii in the \emph{xy}-plane (compare Figs. 2, 3
and 4), \emph{i.e.}, for the magnetic field applied along the
\emph{z-}direction an increase of the absolute value of the
Coulomb energy for both electron-heavy-hole and
electron-light-hole pairs is observed. Only a very small increase
with applied magnetic field along the plane of the dot (see the
dashed and dotted curves with circles and squares in Fig. 4) is
seen.

\section{Magneto-exciton transitions and comparison between theory and experiment}

\begin{figure}
\vspace{-0.5cm} \ \epsfig{file=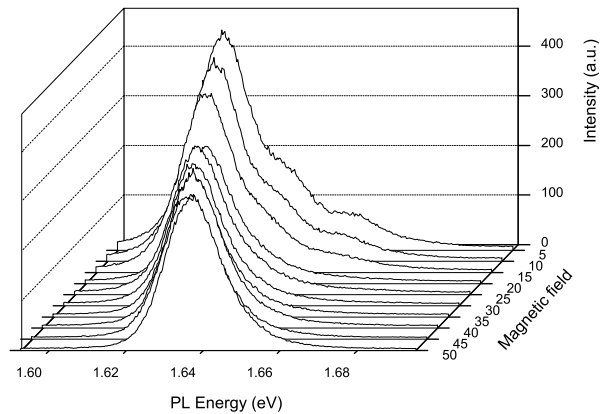,width=9.5cm}
\vspace{-0.8cm} \ \caption{ Photoluminescence spectra as a
function of magnetic field applied parallel  to the
\emph{z-}direction. The peak positions of ground and excited
states are obtained by gaussian fits.}
\end{figure}

In this section we investigate the exciton diamagnetic shift of
the dot. We perform our simulations for both heavy- and light-hole
states and compare them with the experimental data. The
experimental data are obtained by photoluminescence measurements
in pulsed magnetic fields up to 50 Tesla (see Fig. 5). The
experiments were carried out in a He$^{4}$-bath cryostat at 4.2 K,
using an argon-ion laser at a wavelength of 514 nm to excite the
electrons. The photoluminescence (PL) was analysed by a
spectrometer and an intensified charged-coupled-device detector,
using an integration time of 0.5 ms. The peak positions of the PL
spectrum are obtained by gaussian fits to the experimental
spectra. The measured PL ground state energy versus field
dependence can be fitted with an excitonic model, using three
parameters: the zero field PL energy, exciton wave function radius
and the exciton mass. The exciton radius in the \emph{xy-}plane
obtained with the fitting procedure from the experimental data of
7.6 nm is in good agreement with the calculated electron radius as
described in the previous section. A detailed description of the
experimental set-up and analysing method can be found in
Ref.~\onlinecite{Leuven}.

\begin{figure}
\vspace{-0.5cm} \ \epsfig{file=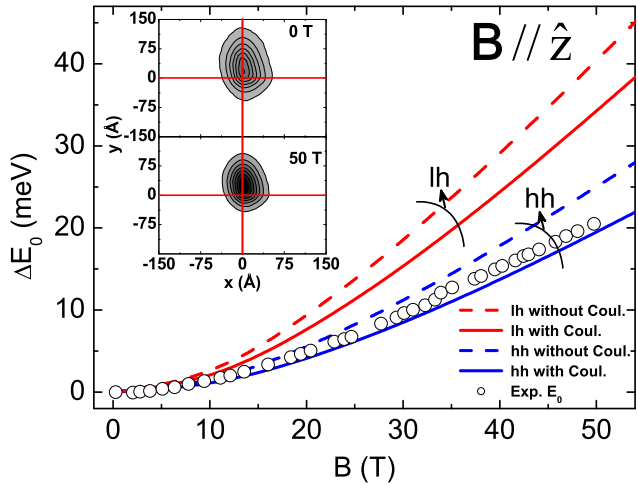,width=9.5cm}
\vspace{-0.8cm} \ \caption{(color online). The diamagnetic shift
of the exciton ground state energy, as a function of a magnetic
field along the \emph{z-}direction for GaAs/Al$_{x}$Ga$_{1-x}$As
QD. The full and dashed curves correspond to the heavy-hole and
the light-hole excitons with and without taking into account the
Coulomb interaction. The inset shows the contour plot of the
electron density at magnetic field 0 and 50 T.}
\end{figure}

In Fig. 6 we examine the diamagnetic shift of the ground state
exciton energy, with and without taking into account the Coulomb
interaction, when the magnetic field is applied along the
\emph{z-}direction. From Fig. 6 we can immediately notice the
importance of including the Coulomb interaction energy (compare
full and dashed curves). The resulting diamagnetic shift is
inversely proportional to the reduced mass of the electron and
hole. The heavy-hole mass is four times larger in the
\emph{xy}-plane of the dot than the light-hole. Consequently, the
resulting heavy-hole diamagnetic shift is smaller than the light-
hole diamagnetic shift, when the magnetic field is applied along
the growth direction (\textbf{B}$//\hat{z}$). Up to 35 T the
calculated diamagnetic shift energies for the heavy-hole states
are in very good agreement with the experimental data. At 50 T the
disagreement between theory and experiment is less than 1 meV. If
we exclude the nonparabolicity of the electron mass the
disagreement between the theoretical heavy-hole exciton and the
experiment at 50 T is 3 meV. In the inset of Fig. 6 we show a
contourplot of the electron density which illustrates the
asymmetry of the electron density regarding the \emph{xy}-center.
The zero magnetic field PL energies for the heavy-hole state is
E$_{PLhh}$=1614 meV, which compares to the experimental PL energy
E$_{PLExp.}$=1620 meV, while for the light-hole we found
E$_{PLlh}$=1652 meV. We neglected the Zeeman effect in our
calculations, since even at the highest magnetic field it gives
only a small contribution. Taking the g$_{ex}$-factor for the
exciton equal to 0.51~\cite{g factor}, the Zeeman splitting energy
between the states with the spin up and down is only
g$_{ex}$$\mu$$_{B}$B $\approx$ 1.5 meV at 50 T, where $\mu$$_{B}$
is the Bohr magneton.

\begin{figure}
\vspace{0.5cm} \ \epsfig{file=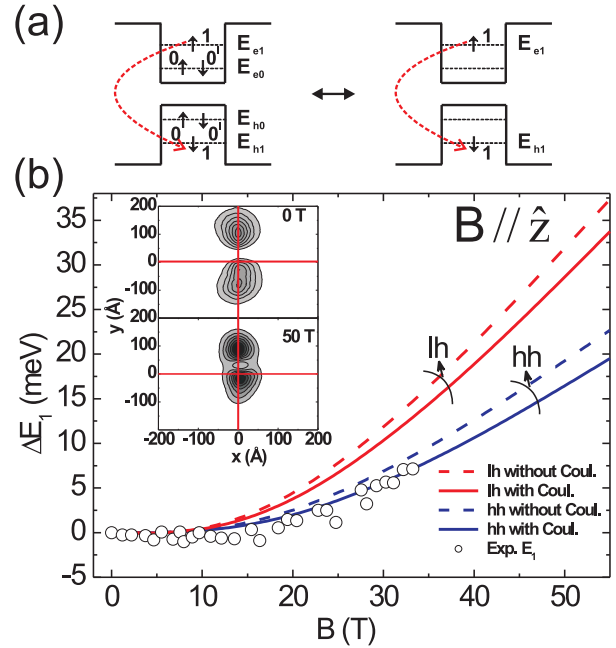,width=9.5cm}
\vspace{-0.8cm} \ \caption{(color online). (a) Schematic of the
excitonic levels for the first excited state. (b) The diamagnetic
shift of the first excited exciton state, as a function of
magnetic field applied along the \emph{z-}direction for
GaAs/Al$_{x}$Ga$_{1-x}$As QD. The full and dashed curves
correspond to the heavy-hole and the light-hole excitons with and
without taking into account the Coulomb interaction. The inset
shows the contour plot of the electron density of the first
excited state at magnetic field 0 and 50 T.}
\end{figure}

Figure 7(b) shows the exciton diamagnetic shift of the first
excited state in the presence of a magnetic field applied along
the \emph{z-}direction. In the experiment, the PL intensity of the
first, and particulary the second, excited states is quenched with
magnetic field, such that experimental data is only available up
to 35 and 10 T for the first and second excited states
respectively. Our calculations demonstrate that there is an
enhancement of the electron-hole wave-function overlap with
magnetic field (not shown). Thus, since the intensity of the
ground state PL also becomes slightly less intense with magnetic
field, this implies that electron-hole pair capture by the dots is
suppressed at high fields. Theoretically, we investigate both
heavy- and light-hole exciton states with and without taking into
account the electron-hole interaction. In the whole magnetic field
region of available experimental data, up to 35 T, we again see a
quantitative agreement with the heavy-hole diamagnetic shift when
including the Coulomb interaction. Comparing the PL energies, the
heavy-hole state is also closer to the experimental PL energy
E$_{PLExp.}$=1638 meV, E$_{PLhh}$=1630 meV and E$_{PLlh}$=1673
meV. Furthermore, if we compare the heavy-hole PL energy
difference between the first excited state and the ground state
$\triangle$E$_{PLhh(1,0)}$=16 meV with the experimental values
$\triangle$E$_{PLExp.(1,0)}$=18 meV, it gives further support to
our claim that the heavy-hole exciton is the ground state. In the
inset of Fig. 7(b) we illustrate the electron densities of the
first excited state for a magnetic field of 0 and 50 T.

The exciton energy was calculated, as E$_{ex,i}$ = E$_{e,i}$ +
E$_{h,i}$ - E$_{C(ei,hi)}$, where \emph{i} = 0 refers to the
ground state, \emph{i} = 1 to the first excited state, and so on;
and E$_{C(ei,hi)}$ is the Coulomb interaction between the
electron-hole (heavy- and light-holes) pair in the \emph{i}-th
state. Knowing this exciton energy, we calculate the exciton
diamagnetic shift energy $\bigtriangleup$E$_{i}$. One should
stress, that we assume the recombination between electron and hole
of the same quantum state, namely, electron and hole are either
both in the ground state or in the first or second excited state,
which is experimentally and theoretically
justified~\cite{ExciteB1,ExciteB2,ExciteB1a,ExciteB2a} on the
basis of selection rules.
\begin{figure}
\vspace{-0.5cm} \ \epsfig{file=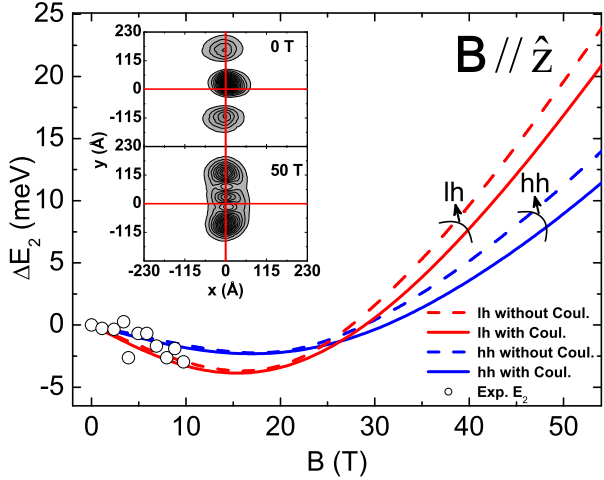,width=9.5cm}
\vspace{-0.8cm} \ \caption{(color online). The same as Fig. 6 but
now for the second excited state.}
\end{figure}
Moreover, we consider the possibility of multiexcitonic
transitions. Let us compare two situations (1) each particle is in
the first excited state (see the right graph in Fig. 7(a)), which
corresponds to the first excited state of the exciton, and (2) the
two identical particles (electron or hole) with spin up and down
are in the ground state and one in the first excited state (see
the left graph in Fig. 7(a)), which corresponds to the ground
state of the three exciton-line. In the first case we have for the
diamagnetic shift for the excited exciton the following terms
$\triangle$E$_{e1}$ + $\triangle$E$_{h1}$ - E$_{C(e1,h1)}$. In the
second case, instead of -E$_{C(e1,h,1)}$ appears 2E$_{C(e0,e1)}$ +
2E$_{C(h0,h1)}$ - 2E$_{C(e0,h1)}$ - 2E$_{C(e1,h0)}$ -
E$_{C(e1,h1)}$. Here E$_{C(e0,e1)}$ is the electron correlation
energy of the ground and first excited states, E$_{C(h0,h1)}$ is
the hole correlation energy, and the rest are the Coulomb
interaction energies between the ground and first excited states
of the particles. We computed the aforementioned correlation and
Coulomb interaction energies for the heavy-hole state, as an
example, in order to estimate the distinction. However, we found
that the difference between the first excited state of the single
exciton transition and the ground state of the three exciton-line
transition is very small. It is less than 1 meV at zero magnetic
field, and about 1 meV at 50 T. The same situation arises if we
compare the ground state of the exciton (electron and hole are in
the ground state) and the ground state of the two exciton-line
(two electrons and holes are in the ground state). The difference
is also less than 1 meV at magnetic field 0 and 50 T.

\begin{figure}
\vspace{-0.5cm} \ \epsfig{file=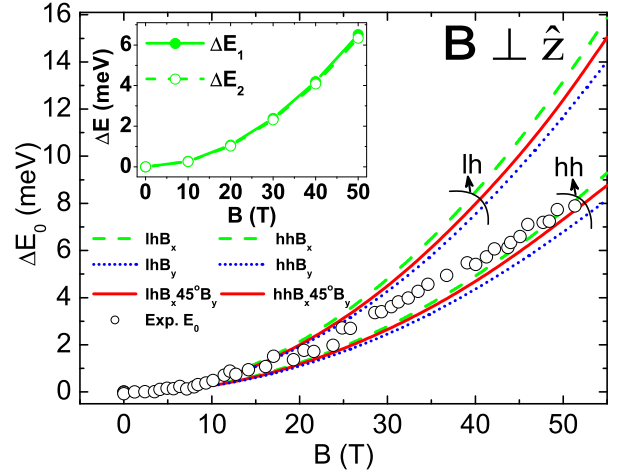,width=9.5cm}
\vspace{-0.8cm} \ \caption{(color online). The diamagnetic shift
of the exciton ground state, as a function of magnetic field. The
dashed (green) and dotted (blue) curves correspond to the heavy-
hole and the light-hole excitons, when the magnetic field is
applied along the \emph{x-} and \emph{y-}direction, respectively.
The full (red) curves correspond to the case, when the magnetic
field is applied along 45$^{\circ}$ between \emph{xy}-direction.
The inset shows the heavy-hole exciton diamagnetic shift of the
first (E$_{1}$) and second (E$_{2}$) excited state, when the
magnetic field is applied along the \emph{x-}direction.}
\end{figure}

Comparison between our theoretical simulations for the exciton
diamagnetic shift of the second excited state and available
experimental data are shown in Fig. 8. Since there are only
experimental data for magnetic fields up to 10 T, it is difficult
to judge, whether heavy- or light-hole exciton states are in good
agreement with the experiment (see curves for the heavy-hole and
the light-hole with and without Coulomb interaction). Comparing
the PL energies, the heavy-hole state is closer to the
experimental PL energy E$_{PLExp.}$=1650 meV, E$_{PLhh}$=1644 meV
and E$_{PLlh}$=1692 meV. When we compare the heavy-hole PL energy
difference between the second excited state and the first excited
state with the experimental data, we found that
$\triangle$E$_{PLhh(2,1)}$=14 meV and
$\triangle$E$_{PLExp.(2,1)}$=12 meV, which is another
justification that the experimental results correspond with
heavy-hole excitons. However, for both hole states we observe a
negative diamagnetic shift, with increasing magnetic field for B
$<$ 15 T, which agrees with the experimental data (see open
circles). Let us remember, that including the magnetic field
causes an extra parabolic confinement, as described by Eq. (7).
The well known Fock-Darwin spectrum at small values of the
magnetic field, where the 2D harmonic oscillator is considered,
shows a small negative dependence of the first excited state
energy, which is more pronounced after the crossing of the second
and third excited state energies. However, the negative excitonic
diamagnetic shift is seen for the second excited state both in the
theoretical and experimental results (compare Figs. 7 and 8). In
the inset of Fig. 8 the change of the electron density for the
second excited state for the magnetic fields of 0 and 50 T is
plotted.

We also investigate the exciton diamagnetic shift of the QD in the
presence of a magnetic field perpendicular to the
\emph{z-}direction (\textbf{B}$\perp\hat{z}$). As an example we
consider the exciton diamagnetic shift, when the magnetic field is
applied along the \emph{x-}, \emph{y-}direction, and directed
45$^{\circ}$ between the \emph{xy}-direction (see Fig. 9). The
Coulomb interaction energy, as was already discussed in Fig. 4,
has a negligible effect on the magnetic field dependence. For
magnetic fields up to 40 T the experimental points are between the
heavy- and light-hole curves, and for high magnetic field a good
agreement between the experimental points and the theoretical
heavy-hole diamagnetic shift is obtained. Within the single band
model it is not possible to take into account the coupling between
the heavy-hole and the light-hole, which is the most likely reason
for the discrepancy at magnetic fields up to 40 T. Note, that
there is a large distinction in the diamagnetic field between
heavy- and light-hole states. The difference for the magnetic
field applied along the \emph{x-}, \emph{y-}direction, or
45$^{\circ}$ between \emph{xy}-direction is small. The inset of
Fig. 9 shows the heavy-hole exciton diamagnetic shifts (the
light-hole excitons have the same behavior) for the two lowest
excited states. The dot confinement is more important than the
magnetic confinement at small magnetic field. Hence, we do not
observe a negative diamagnetic shift behavior for both excited
states.

\section{Conclusions}

In summary, the excitonic properties in the presence of a magnetic
field in unstrained GaAs/Al$_{x}$Ga$_{1-x}$As QDs were
investigated theoretically and experimentally. In our 3D
calculations, within the single band effective mass approximation,
we include the effect of band nonparabolicity of the conduction
band, the real shape of the QD with the thin well layer on top and
asymmetrical barrier materials, as well as heavy- and light-hole
states. By solving the Poisson equation with the correct boundary
conditions, the Coulomb interaction energy between electron and
hole was calculated.

The magnetic field dependence of the electron-hole Coulomb
interaction energy, when the magnetic field is applied along
\emph{x-}, \emph{y-} and \emph{z-}direction, follows closely the
dependencies of the particle radii (\emph{i.e.} size) in the
\emph{xy}-plane of the dot. Radii along the \emph{z-}direction are
significantly smaller due to the flatness of the dot, and with
increasing magnetic field along the \emph{z-}direction the
particles wave function radii become elongated in the
\emph{z-}direction.

The diamagnetic shift for the exciton ground state and the two
first excited states in the presence of a magnetic field applied
along \emph{x-}, \emph{y-}, and \emph{z-}direction in the QD are
calculated. A good agreement between the experimental results and
the calculated heavy-hole exciton is found. By comparing the PL
energies, the heavy-hole state is found to be the ground state,
and its value is close to the experimental data.

\section{Acknowledgments}

This work was supported by the Belgian Science Policy (IUAP), the
European Commission network of excellence: SANDiE Contract No.
NMP4-CT-2004-500101 and EuroMagNet Contract No.
RII3-CT-2004-506239. The authors acknowledge fruitful discussions
with A. Schliwa, C. Manzano, G. Costantini, and K. Kern.

\end{document}